\documentclass[aps,prb,twocolumn,showpacs,letterpaper,superscriptaddress]{revtex4-1}
\usepackage{graphicx}
\usepackage{epstopdf}
\usepackage{multirow}
\usepackage{amsmath}
\usepackage{color}
\usepackage{hyperref}
\usepackage{makecell}
\usepackage{mathtools}
\usepackage{tabularx}
\usepackage{multirow}
\hypersetup{
    colorlinks=true,
   linkcolor=blue,
    filecolor=blue,
		urlcolor=blue,
   citecolor=blue,
}
\usepackage[normalem]{ulem}

\begin{document}

\title{Temperature-dependent dielectric function of  bulk SrTiO$_3$: Urbach tail, band edges, and excitonic effects}

\author{Pranjal Kumar Gogoi}
\email{pkgogoi@gmail.com}
\affiliation{Singapore Synchrotron Light Source, National University of Singapore, 5 Research Link, 117603, Singapore}
\affiliation{NUSNNI-NanoCore, National University of Singapore,  117576, Singapore}
\affiliation{Department of Physics, Faculty of Science, National University of Singapore,  117542, Singapore}

\author{Daniel Schmidt}
\affiliation{Singapore Synchrotron Light Source, National University of Singapore, 5 Research Link,  117603, Singapore}


\date{\today}
%
%
\begin{abstract}
We report the temperature-dependent complex dielectric function of pristine bulk SrTiO$_3$ between 4.2 and 300 K within the energy range of 0.6-6.5 eV determined by spectroscopic ellipsometry. Fundamental indirect and direct band-gap energies have been extracted and are discussed with regard to existing state-of-the-art theoretical calculations. Furthermore, the dielectric function around the fundamental direct gap is analyzed by considering excitonic states. The excitonic effects, including the Coulomb enhancement of the continuum, are characterized using an extension of the Elliott's formula considering both the real and imaginary parts of the dielectric function. The Urbach tail below the indirect edge shows an unconventional temperature-dependent behavior correlated to the microstructural changes near the structural phase transition around 105 K from the low-temperature tetragonal phase to the cubic phase. The temperature-dependent characterization reveals that the fundamental indirect edge as well as the Urbach tail are affected conspicuously by the structural phase transition while the fundamental direct edge is not. Moreover, the indirect edge follows Varshni's rule only in the cubic phase and the direct edge exhibits an anomalous linear increase with increasing temperature.
\end{abstract}


\maketitle

\section{\label{intro}Introduction}

Despite being a very widely used and studied material for more than half a century now, SrTiO$_3$ (STO) still continues to intrigue physicists and material scientists alike. The usefulness and importance of STO, for example, are highlighted by its numerous technological applications in various forms~\cite{Dawber, Sawa, beck} and its role as an ideal substrate  for thin-film growth, particularly for high-temperature superconducting films~\cite{Dijkkamp}. From a fundamental physics perspective, for example, it is used as a model prototype perovskite transition metal oxide for theoretical as well as  experimental  studies of many-body correlated phenomena~\cite{Sponza, Gogoi}, and structural~\cite{Mueller1} as well as quantum phase transitions~\cite{Mueller3, rowley}.  The relevance  of STO in current  science and near future technology has been further enhanced by its recent emergence as an integral  component of  oxide interfaces with exotic properties~\cite{Kawasaki, Hwang, Kozuka, Asmara,Reyren, Li, Bert, Brinkman, Ariando, Ohtomo, Thiel}. Conducting interfaces between two insulating  oxides, where STO is one of the most common components, show tunable quasi-two-dimensional electron gas characteristics, superconductivity, metal-insulator transition, and other rich phenomena which thrive on key fundamental principles of modern condensed matter physics. Remarkably, recent studies have revealed that the bare STO surface itself shows many of these intriguing properties~\cite {santander2011, Meevasana, santander2014}. Similarly, bulk STO  with  intrinsic defects~\cite{szot}, with doping~\cite{watanabe, vanmechelen}, after reduction~\cite{Schooley}, as well as thin-film STO with strain~\cite{haeni}, show exquisite properties originating from complex and fascinating interactions of the quasiparticles.

Hence, with its  prominence as a key material, a thorough understanding of STO in terms of the underlying physics is crucial from the point of view of both fundamental sciences as well as applications. The investigation of various aspects of  the electronic structure in general is an important step towards achieving that. Optical spectroscopy techniques and particularly spectroscopic ellipsometry can be used to determine the  dielectric function, which provides valuable information about the electronic structure of the material. Not surprisingly, there have been numerous experimental  studies on STO in the past  with the aim of elucidating the electronic structure employing various types of optical spectroscopies~\cite{Noland, Cardona, Cohen, Capizzi, Blazey, RedfieldPRL, Redfield, Bauerle, Servoin, Goldschmidt, Jellison,  Zollner, Hasegawa, vanBenthem, Trepakov, Yamada, Yao,  Lee, Dorywalski,  Kok}. Simultaneously, many theoretical studies have been performed on STO using different approaches~\cite{Kahn, Matthiess,  Ahuja, Piskunov, Gupta, Samantaray}. However, only two include electron-hole ($e$-$h$) interactions, which turn out to be crucial for understanding the optical spectra~\cite{Sponza, Gogoi}. The majority of reports agree that  STO has an indirect edge  around 3.2 eV and a direct edge  below 4 eV. However, in spite of  the relatively large amount of experimental and theoretical works, still there have been considerable disagreements on the exact position of the band edges and particularly on their origins.   At the same time, the  identification of critical points of the dielectric function and  their relationship to the optical transitions in the band-structure have been ambiguous~\cite{Cardona, Bauerle, vanBenthem, Zollner}. 

A proper understanding of the optical spectra in terms of the electronic structure is possible by the comparison of low-temperature experimental optical spectra with state-of-the-art \textit{ab initio} calculations of the dielectric function~\cite{Albrecht, Rohlfing, CardonaAspnes, AlbrechtReply}. The first reports on low-temperature optical studies on STO in the early 1970s focused on the identification and temperature evolution of the fundamental indirect edge~\cite{Capizzi, Blazey, RedfieldPRL, Redfield}. However, the assignment of optical transitions as well as the involved phonons are not in agreement anymore with the advanced theoretical calculations. More recently, several studies investigated the indirect edge at low temperatures with photoluminescence and absorption measurements~\cite{Hasegawa,Kok,Yamada}. Trepakov \textit{et al.}~\cite{Trepakov} employed spectroscopic ellipsometry to characterize the temperature dependence of the indirect as well as the direct edge~\cite{Trepakov}. However, their characterization only includes temperatures down to 110~K and excitonic influences on the direct edge have not been considered. Recently, we have reported the observation of excitons in STO and Nb-doped STO based on temperature-dependent changes in the pseudodielectric function in conjunction with \textit{ab initio} calculations~\cite{Gogoi}. However, there was no focus on the detailed characterization of the temperature evolution of the absorption tail and critical points.


In this work, we report the complex dielectric function of STO from room temperature down to 4.2 K using spectroscopic ellipsometry in the energy range from 0.6 to 6.5~eV. This wide spectral range allows characterization of the critical points of interest, involving transitions from various upper valence band states to low-lying conduction band states as a function of temperature across the structural phase transition at around 105~K. Critical point identification based on recent \textit{ab initio} \textit{GW} Bethe-Salpeter equation calculations as well as thorough line-shape analysis of the dielectric function have been performed. Both the fundamental indirect and direct edges are investigated in detail for the full temperature range and are brought into a coherent picture. The temperature dependence of the indirect edge is found to follow Varshni's rule down to the structural phase transition. The fundamental direct edge has been estimated by taking $e$-$h$ interactions into consideration, in contrast to the previous reports where a noninteracting direct band edge was assumed. The temperature dependence of the direct gap exhibits an anomalous characteristic by linearly increasing with increasing temperature.

We would like to point out that, while investigating various previous reports on the dielectric function of STO, with particular focus on the apparent inconsistencies, it is found that an unfortunate error led to further augmentation of  the uncertainties. While converting previously reported dielectric function data from Cardona and others~\cite{Cardona, Bauerle} to refractive index and extinction coefficient, apparently the data was redshifted by $\sim$200~meV~\cite{Gervais}. This mistake was then repeatedly reported by various later authors  along with their new results, each explicitly highlighting these discrepancies~\cite{vanBenthem, Zollner, Dorywalski, Sponza}. We find that the original data by Cardona in fact agree quite well with most subsequent reports.



\begin{figure}
\includegraphics[width =\columnwidth, clip, trim=0 0 0 0 ]{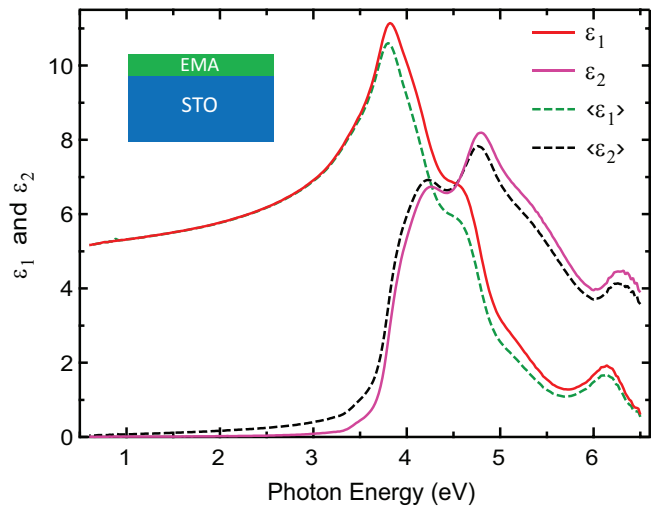}
\caption{\label{fig1} 
Dielectric function $\varepsilon$ and pseudodielectric function $\langle\varepsilon\rangle$ of STO at 300 K. The inset shows the optical model with surface layer (EMA) on top of the STO substrate.}
\end{figure}

\section{\label{exp}Spectroscopic Ellipsometry and Dielectric function extraction}

In spectroscopic ellipsometry, the polarization state change of light upon reflection from (or transmission through) a sample is measured. The fundamental ellipsometric relationship is expressed as~\cite{azzam, Fujiwara}
\begin{equation} \label{ellips}
\varrho = {r_p}/{r_s} = \tan\Psi e^{i\Delta},
\end{equation}

\noindent where the  ellipsometric angles $\Psi$ and $\Delta$ are  the measured quantities  and $r_p$   and $r_s$   are the Fresnel reflection coefficients for $p$ and $s$ polarized light, respectively.  The Fresnel reflection coefficients are related to the complex dielectric function, $\varepsilon = \varepsilon_1 + \mathbf{\rm i} \varepsilon_2$, of the sample~\cite{Fujiwara}.

Spectroscopic ellipsometry measurements are  performed in the spectral range from  0.6 to 6.5 eV  with a rotating analyzer ellipsometer with compensator (V-VASE, J.A. Woollam Co., Inc.). All measurements are performed inside  a cryostat (Janis) at ultrahigh vacuum with  base pressure in the  10$^{-9}$ torr regime.  The angle of incidence is kept at  $70^\circ$ for all the measurements.  Measurements are performed from 4.2  to 300 K in 25 K intervals using liquid helium and liquid nitrogen as  cryogens  in an open cycle configuration.

Single side polished SrTiO$_3$ (100)  samples of size 10 $\times$ 10 $\times$ 0.5  mm$^3$, procured from CrysTec GmbH, were used for the measurements.  Atomic force microscopy measurements show that the rms roughness is less than 5 \AA.


In the case of isotropic bulk samples, in theory the dielectric function can be directly calculated from the measured ellipsometric quantities $\Psi$ and $\Delta$. However, in practice, this dielectric function usually has contributions from surface effects such as surface roughness and/or adsorbed overlayers~\cite{Nelson} and is therefore called the pseudodielectric function $\langle\varepsilon\rangle$. 
These contributions cause, among other differences, a pseudoabsorption ($\varepsilon_2 > 0$) far below the first optical transition. In order to describe these surface effects, typically a practical approach is to use a Bruggeman effective medium approximation (EMA) assuming 50\% bulk and 50\% void~\cite{Nelson, Tompkins1999}. 

To find the thickness of the surface layer in the initial step, data in the transparent region below 2.3 eV are fitted using a Cauchy dispersion with zero absorption~\cite{Fujiwara}. A three-phase structure with ambient/surface layer/sample is employed as the optical model, where the surface layer is represented by the Bruggeman effective medium layer. This surface layer thickness is then used for the wavelength-by-wavelength fit of the ellipsometric data for the full energy range using a mathematical inversion approach, i.e., the only unknown parameters $\varepsilon_1$ and $\varepsilon_2$ can be calculated from the measured ellipsometric angles~\cite{Fujiwara}. It is to be noted that a parameterized fit is not used for extraction of the dielectric function, since using any functional oscillator models may result in smoothing out of spectral features~\cite{schmidt2015}.

In Fig.~\ref{fig1}, $\langle\varepsilon\rangle$  is plotted for STO at 300~K along with the extracted dielectric function. The thickness of the model surface layer is found to be 1.7~nm from the initial Cauchy fitting.  Note that measurements outside the cryostat in ambient conditions result in a thickness of 2.1~nm. These values are consistent with previous ellipsometric studies on STO and also with the fact that the adsorbed overlayers decrease in ultrahigh vacuum conditions due to desorption~\cite{Zollner, Dorywalski}.

\begin{figure}
\centering
\includegraphics[width =\columnwidth, clip, trim=35 45 20 0] {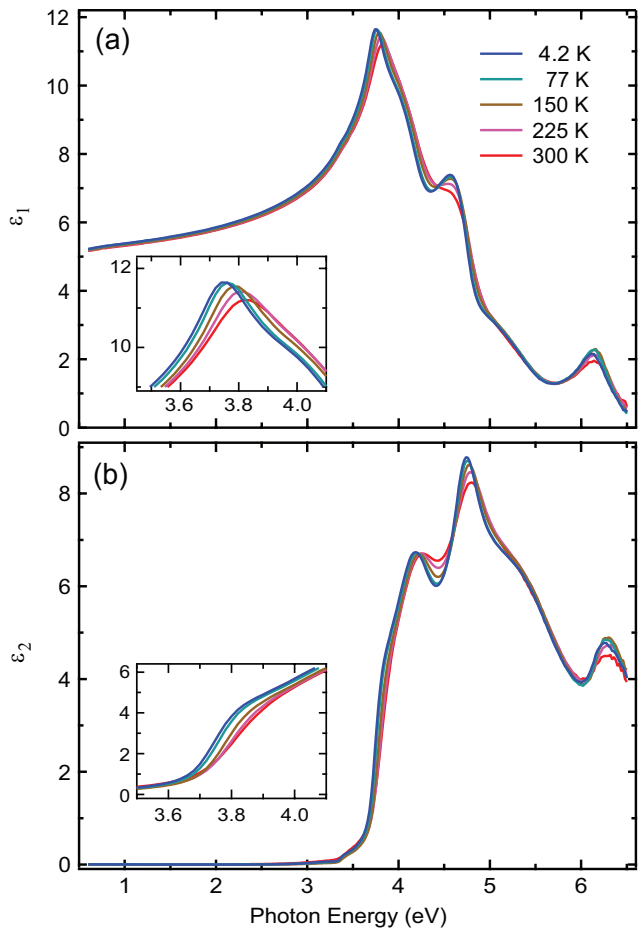}
\caption{\label{fig2}
Temperature dependence of the  real (a) and imaginary part (b) of the dielectric function of STO~\cite{suppl}. The insets highlight the fundamental direct edge (around 3.8 eV).}
\end{figure}

\section{\label{results} RESULTS AND DISCUSSION}

\subsection{\label{subs_DF} Dielectric functions from 4.2 K upto  300 K}

Real ($\varepsilon_1$) and imaginary ($\varepsilon_2$) parts of the dielectric function at selected temperatures are shown in Figs.~\ref{fig2}(a) and \ref{fig2}(b), respectively. As the dielectric function is directly related to the band-structure and the electronic correlations, we base our interpretation of the various features  on reported state-of-the-art band-structure and optical spectra calculations of STO~\cite{Sponza, Piskunov}. In STO, crystal field effects break the symmetry of both the 3$d$ and 4$d$ orbitals of Ti and Sr, respectively. The fivefold degenerate $d$ levels get split into two subgroups $t_{2g}$ ($d_{xy}$, $d_{yz}$, $d_{zx}$) and $e_{g}$ ($d_{z^2}$ and $d_{x^2-y^2}$). The lowest conduction band states comprise Ti 3$d$-$t_{2g}$ orbitals and are at the $\Gamma$ point. These Ti 3$d$-$t_{2g}$ bands span an energy range of about 2.6 eV above the conduction band minimum within the Brillouin zone. The Ti 3$d$-$e_{g}$ and Sr 4$d$-$e_{g}$ levels together comprise the next group of bands starting from around this point. However, between these two, in our energy range of interest, only levels originating from Ti 3$d$-$e_{g}$ orbitals take part in the relevant transitions \cite{Sponza, Piskunov}.

The valence band essentially comprises  nine bands originating from O 2$p$ orbitals, which are in a hybridized state with Ti~3$d$-$t_{2g}$ orbitals. At the $\Gamma$ point, these bands are occurring in three subgroups, and each one is threefold degenerate. The top of the  valence band is at the $R$ point in the Brillouin zone~\cite{Sponza, Piskunov}.

Since no lattice vibrations were assumed in the theoretical calculations considered here~\cite{Sponza, Piskunov}, it is important that comparisons of theoretical results are made with low-temperature experimental results~\cite{CardonaAspnes}. Although the general features of the dielectric function for all temperatures are similar, for low temperatures shifts and a distinct sharpening of peaks as well as an increasing prominence of other structures are seen. An interesting difference between the room temperature and the 4.2-K dielectric function is the distinct structure at around 3.8 eV observed for the latter. This structure is barely discernible in the room temperature dielectric function and hence  was ignored before in previous room temperature studies~\cite{Cardona, vanBenthem}. However, it has very important ramifications in the understanding of the properties of STO,  as we find that it represents a three-dimensional $M_0$ critical point with excitonic enhancement~\cite{YuCardona, Mcardona} (see also Sec. \ref{subs_DE}).

With the help of reported state-of-the-art band-structure calculations, we assign various features (e.g., peaks and shoulders) in the $\varepsilon_2$ to transitions occurring at high-symmetry points in the Brillouin zone~\cite{Sponza, Piskunov}. In Table~\ref{tab1} these assignments are presented together with the  \emph{GW} gap value, which is related to the peak or structure in $\varepsilon_2$, estimated from the band-structure calculation of Sponza \textit{et al.} \cite{Sponza}. It should be noted that the exact gap energy is not necessarily corresponding to the peak energy in $\varepsilon_2$. Rather line-shape analyses are required to extract the exact threshold energy of each critical point from the dielectric function~\cite{Lautenschlager}. Nevertheless, here, the peak energies agree quite well with the respective \emph{GW} gap energies. These assignments, discussed in detail below, are based on our 4.2-K data and state-of-the-art theoretical results and eliminate the inconsistencies found in previous reports~\cite{Cardona, vanBenthem, Bauerle, Zollner, Piskunov}.

\begin{center}
\begin{table}[tbp]
\caption{Identification of optical transitions responsible for the features in $\varepsilon_2$ for STO.\\[-6pt]} \label{tab1}
\renewcommand{\arraystretch}{1.2}
\begin{center}
\begin{tabular*} {0.9\linewidth}{@{\extracolsep{\fill}}cccrrrl} \hline\hline \\[-10pt]
\textbf{\makecell{Features \\in $\varepsilon_2$ (eV)}}     & \textbf{\makecell{\emph{GW} gap\\ (eV)}} & \multicolumn{5}{l}{\textbf{\makecell{Associated high-\\symmetry points}}}\\

\hline
    3.2           &	3.4      &  & & $R$         & $\rightarrow$  & $\Gamma$\footnote{Indirect transition.}  \\
	3.8             &	3.8      &  & & $\Gamma$  & $\rightarrow$  & $\Gamma$ \\
	4.2             &	4.4      &  & & $X$         & $\rightarrow$  & $X$ \\
  4.8             &	4.8      &  & & $\Gamma$  & $\rightarrow$  & $\Gamma$\\
	5.2             &	5.4      &  & & $M$         & $\rightarrow$  & $M$ \\
  6.0             &	6.0      &  & & $R$         & $\rightarrow$  & $R$ \\
  6.2             &	6.4      &  & & $\Gamma$  & $\rightarrow$  & {$\Gamma$}  \\
\hline

\end{tabular*}
\end{center}
\end{table}
\end{center}

As can be seen in Fig.~\ref{fig2}(b), the absorption onset is near 3.2 eV, which has been attributed to the indirect edge in STO \cite{Capizzi, vanBenthem, Yamada, Hasegawa}. Early theoretical studies employing semiempirical pseudopotential calculations attributed this indirect edge to transitions from the $\Gamma$ point (highest valence band)  to the $X$ point (lowest conduction band) in the Brillouin zone \cite{Kahn, Matthiess, Mattheiss1}. Most experimental works thereafter  used  this theoretical understanding for interpreting their results. However, more recently it was found that the transitions are in fact from the $R$ point in the valence band  to the $\Gamma$ point in the conduction band, where they represent the highest and the lowest points in the bands, respectively \cite{Ahuja, Gupta, Samantaray, Piskunov, Zollner}. In summary, until now, there has been no unambiguous assignment of the indirect gap, which is consistent with both theory and experimental data. Even the most recent results on the low-temperature indirect edge energy have been based on the assignment of the indirect transition from the $\Gamma$ point in valence band to the $X$ point in the conduction band \cite{ Yamada, Hasegawa}. Similarly, other reports in the past have not comprehensively described the phonons involved~\cite{vanBenthem, Kok, Trepakov}. In this study, the fundamental indirect edge has been extracted and is found to be  3.199 eV at 4.2 K.  The details are given in Sec. \ref{subs_IE}.


The next higher energy structure is associated with the fundamental direct edge around 3.8 eV \cite{vanBenthem, Trepakov, Capizzi} and is represented by a prominent shoulder in the 4.2-K dielectric function. It is associated with transitions in the $\Gamma$ point of the Brillouin zone. We find that the dielectric function is modified considerably around the band edge due to excitonic effects, which have to be taken into consideration. A value of 3.78 eV can be estimated for the fundamental direct gap at 4.2 K. The direct edge is discussed in detail in Sec. \ref{subs_DE}.


The peak at around 4.2 eV can be associated with a transition at the $X$ point and the  corresponding \emph{GW} gap is 4.4 eV. As shown by Piskunov \textit{et al.}~\cite{Piskunov}, this is a generic transition of perovskites, with examples given for SrTiO$_3$, BaTiO$_3$, and PbTiO$_3$. The peak in $\varepsilon_2$ at 4.8 eV is the transition at the $\Gamma$ point from the next lower subgroup of the O 2$p$ bands in the valence band to the Ti $3d$-$t_{2g}$ conduction bands. It is interesting to note that the \emph{GW} gap between the valence band maximum at the $\Gamma$ point  and this subgroup of degenerate bands (three bands) is about 1.0 eV. Hence, it is more likely that the 3.8-eV structure at the fundamental direct edge and the 4.8-eV peak are in fact in a doublet relationship, in contrast to  what has been reported earlier~\cite{Cardona, vanBenthem}. The structure at around 5.2 eV can be associated with the direct gap at the $M$ point. The bands are seen to be less dispersive in nature and hence a broad peak structure is seen. Similarly, the subtle 6.0-eV structure  can be associated with the transition at the $R$ point. Our analysis as  explained in Sec.~\ref{subs_DE} can be used as a guide for this identification.

On the other hand, the 6.2-eV  peak is associated with transitions at the $\Gamma$ point. However, since both the O 2$p$ valence and  Ti 3$d$-$e_{g}$ conduction bands are flat along the  $\Gamma$$X$  direction, the peak has contributions from transitions  at these \emph{k} points,  as reported by Sponza \textit{et al.}~\cite{Sponza}. Note that the theoretically calculated large excitonic peak at around 6.3~eV is observed to be significantly less prominent in the experiment even at 4.2~K.


\subsection{\label{subs_UT} Unconventional Urbach tail}

\begin{figure}[tbp]
\centering
\includegraphics[width =\columnwidth, clip, trim=2 0 0 0] {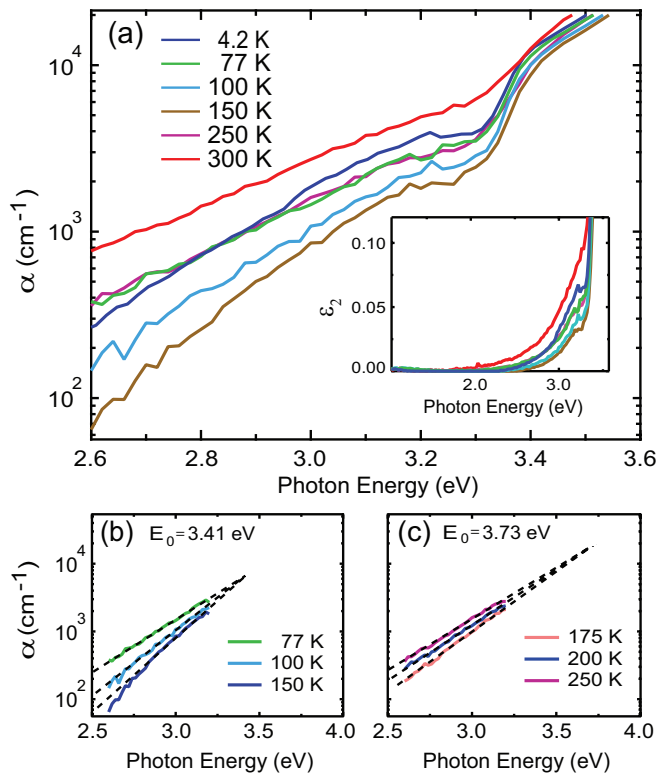}
\caption{Anomalous Urbach tail behavior. (a) Temperature dependence of the absorption coefficient showing the deviation from the Urbach tail behavior of STO near the indirect absorption edge. The inset shows  $\varepsilon_2$ at the same temperatures near the indirect edge. Unconventional (b) and conventional (c) Urbach behavior for the lower- and higher-temperature cases, respectively.}\label{fig3}
\end{figure}

The absorption coefficient $\alpha = \omega\varepsilon_2/cn$, where $c$ is the speed of light, $\omega$ is the angular frequency, and $n$ is the refractive index,  is shown in Fig.~\ref{fig3} for selected temperatures near the fundamental absorption edge ($\sim$3.2 eV). The low-energy tail  ($<$3.2 eV), depicted in Fig.~\ref{fig3}(a),  shows a linear behavior in $\log{(\alpha)}$, which is known as the Urbach tail \cite{Urbach, Kurik}. The Urbach rule is given by
\begin{equation} \label{Urbach}
\alpha (E, T) = {\alpha_0} {e^{- {\sigma}({E_0 - E})/{kT}}},
\end{equation}

\noindent where $E$ is the photon energy, $E_0$ and $\alpha_0$ are constants,  $\sigma$ describes the steepness of the absorption edge (related to the strength of the electron-phonon interactions), $T$ is the temperature, and $k$ the Boltzmann constant. A conventional Urbach tail behavior is characterized by the increasing slope of $\log{(\alpha)}$ vs energy with decreasing temperature. Moreover, the slopes for different temperatures typically converge to an energy value close to the originating transition energy (or gap) when extrapolated. The constant $E_0$ in Eq.~(\ref{Urbach}) refers to this extrapolated convergence point \cite{Urbach, Kurik}.

Remarkably, a deviation from the Urbach tail behavior is observed for the low-temperature spectra of $\alpha$, as shown in Fig.~\ref{fig3}(b). The slopes of $\log{(\alpha)}$ vs photon energy start to decrease below 150 K with decreasing temperature. However, at higher temperatures the conventional Urbach rule is followed, i.e., the slope of the absorption tail  decreases with increasing temperature [Fig.~\ref{fig3}(c)].

Note that the average error bar of the depicted absorption coefficient below 3.2~eV is around 600~cm$^{-1}$. Besides that, changes in the surface layer thickness may offset and alter the slope of $\alpha$ and hence can shift the extrapolated convergence point $E_0$. Nevertheless, changes within the error bar of the determined surface layer thickness ($<2\%$) do not affect the overall anomalous temperature-dependent Urbach tail behavior.  

We propose that this unconventional Urbach tail behavior is a consequence of the crystalline disorder induced by the creation of tetragonal phase grains in the cubic phase near 150 K. It is known that the Urbach tail behavior can also be observed in crystals with microstructural disorder, where the slope of the tail decreases with increasing disorder \cite{Sritrakool, Lifshitz}. Furthermore, recent microstructural studies have shown the creation of tetragonal phase regions and their coexistence with the cubic phase with gradually increasing fraction starting from $\sim$150 K down to $\sim$105 K~\cite{Salman, Singh}. At around 105 K (considered the phase transition temperature) the whole crystal becomes tetragonal. Therefore, relative strengths  of the competing effects of increased microstructural disorder and decreasing temperature determines the final Urbach tail behavior. Below about 150 K, disorder plays the dominant role and hence the unconventional Urbach tail is observed.

The extrapolated convergence point of the slopes shows that $E_0$ for the lower-temperature case is 3.41 eV [Fig.~\ref{fig3}(b)]. This is consistent with previous reports that the corresponding structure in the absorption spectrum is due to the saturation of the indirect transitions \cite{Capizzi, Redfield}. However, for higher temperatures  $E_0$ is 3.73 eV [Fig.~\ref{fig3}(c)], which is close to the fundamental direct transition (see also Sec.~\ref{subs_DE}). Thus the two different $E_0$ observed here suggest that the absorption tails have  fundamentally different origins: at low temperatures it is dominated by the indirect transition, while at higher temperatures it is dominated by the first direct transition. 

Moreover, at 4.2 K, the log plot of $\alpha$ is no longer linear in the energy range of concern and further structures are seen. Additional quasiparticle interactions related to the quantum paraelectric phase could contribute to such new structures \cite{Mueller2, Mueller3, rowley}. Similarly, at 300 K  the slope is not converging to 3.73 eV, indicating that it might be dominated by the next stronger direct interband transition around 4.15 eV.


Deviations from the Urbach tail behavior at low temperatures, however, of different nature than observed here, near the fundamental absorption edge have been mentioned before  for STO. Capizzi and Frova~\cite{Capizzi} reported an oscillatory behavior of the Urbach tail at around 3.30 eV at liquid nitrogen temperatures ($\sim$82 K), attributing it to be caused by the fundamental indirect edge. However, the Urbach tail was investigated only in the immediate vicinity of the fundamental indirect edge. Another report by Hasegawa \textit{et al.}~\cite{Hasegawa} mentioned the deviation of the Urbach tail below 160 K without further explanation. Note that deviations from the Urbach tail behavior have also been observed at low temperatures for the case of rutile TiO$_2$, which has the same Ti-O coordination as STO~\cite{Tang, Masao}. The TiO$_6$ octahedra, which is also present in case of rutile TiO$_2$, is distorted in the tetragonal phase of STO~\cite{Mueller1}. This further supports our explanation that microcrystalline disorder induced by the phase transition is responsible for the unconventional Urbach tail behavior. 


\subsection{\label{subs_IE} Temperature dependence of the fundamental indirect gap}

\begin{figure}
\centering
\includegraphics[width =\columnwidth, trim=10 15 0 0] {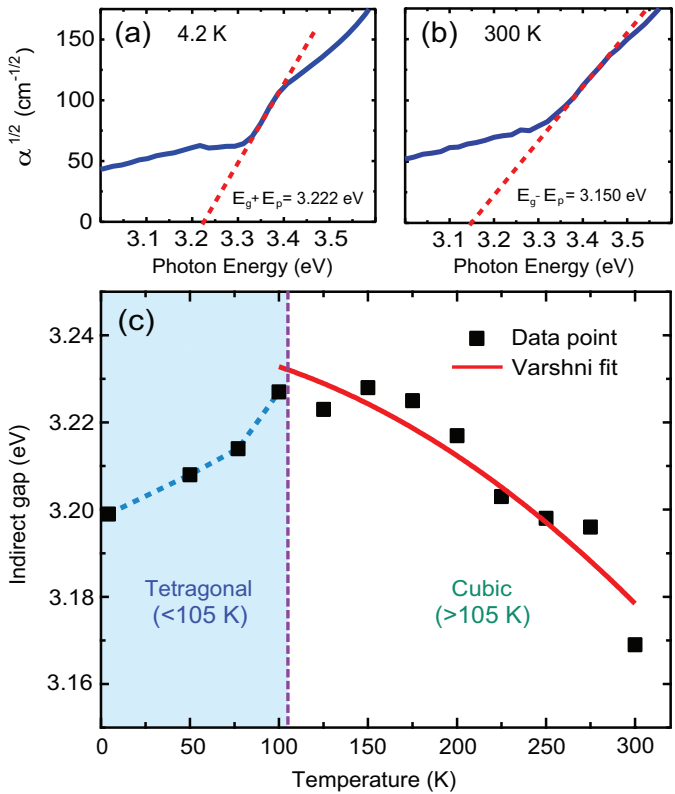}
\caption{Indirect absorption edge  of STO. Linear fits of the square root of the absorption coefficient showing the dominance of the (a) phonon emission at 4.2 K and (b) phonon absorption at 300 K. (c) Temperature dependence of the indirect edge, and  Varshni fit of data from 100 K and above. }\label{fig4}
\end{figure}

 The absorption coefficient near an indirect edge can be described by the following expression:
\begin{equation} \label{Abscoeffind}
\alpha (E) =\frac{A'{(E- E_{\rm g} - E_{\rm p})}^2}{e^{\frac{E_{\rm p}}{kT}} -1} + \frac{A'{(E- E_{\rm g} + E_{\rm p})}^2}{1 - e^{- \frac{E_{\rm p}}{kT}}},
\end{equation}

\noindent where the  first and second term  on the right-hand side represent transitions with phonon absorption and emission, respectively. $A'$ is a constant, which depends on the effective masses of electrons in the conduction band and holes in the valence band. $E_{\rm g}$ is the indirect band-gap energy and $E_{\rm p}$ is the energy of the phonon absorbed or emitted~\cite {YuCardona, Pankove}. Hence, the indirect band-gap energy may be found by extrapolation of the linear regimes in the square root of $\alpha$.  At low temperatures the phonon density is very low and hence the plot of $\alpha^{1/2}$ is dominated by a linear regime representing  phonon emission [Fig.~\ref{fig4}(a)].  Similarly, the phonon absorption  branch dominates the $\alpha^{1/2}$ plot at  high temperatures.

Here, like previous reports~\cite{Trepakov, Kok} and based on our experimental results, we treat the observed linear portions in $\alpha^{1/2}$ to be due to the phonon emission part for temperatures below 225~K and due to the phonon absorption part for temperatures above 225~K. In Figs.~\ref{fig4}(a) and \ref{fig4}(b) the linear fits are shown for $\alpha^{1/2}$ at 4.2 K and 300 K as representative examples of the two different cases,  respectively. Both branches are observed for intermediate temperatures between 175 K and 250 K. For certain cases, more than two linear branches might be observable and can be assigned to impurity-assisted phonon absorption~\cite{Capizzi}.

The indirect transitions have been assigned to be from the $R$ point in the valence band to the $\Gamma$ point in the conduction band based on recent \textit{ab initio} calculation results~\cite{Piskunov, Sponza, Evarestov}. According to previous reports of phonon energies, the phonons responsible for the indirect transitions from the $R$ point are 23- and 18- meV phonons for the tetragonal ($<105$~K)~\cite{Galzerani} and cubic phase ($>105$~K)~\cite{Stirling1972, Servoin}, respectively~\cite{Evarestov, Petzelt}. These assignments are in reasonable agreement with our experimental results and also with the absorption coefficients reported in previous studies, which estimated the fundamental indirect edge~\cite{Yamada, Hasegawa}.  With this scheme,  indirect band gaps are extracted for different temperatures and are plotted with respect to temperature in Fig.~\ref{fig4}(c). Interestingly, the band gap rises slowly until the phase transition temperature starting from 3.199 eV at 4.2 K to 3.226 eV at 100 K and  it decreases gradually thereafter.

Typically band-gaps, both direct and indirect, of most materials follow the empirical  Varshni's equation, which essentially means a nonlinear decrease of the gap $E_{\rm g}$ with increasing temperature given by \cite{Varshni}
\begin{equation} \label{Varshni}
 E_{\rm g} = E_{\rm g_0} -  \frac {\alpha_{\rm V} T^2}{T + \beta_{\rm V}},
\end{equation}

\noindent where $\alpha_{\rm V}$ and $\beta_{\rm V}$ are constants and $E_{\rm g_0}$ represents the energy gap at 0~K. Interestingly, the temperature dependence of the fundamental indirect edge can be described by Varshni's equation only for the cubic phase of STO (above 100~K), as shown in Fig.~\ref{fig4}(c). In this case $E_{\rm g_{0}}$ represents the extrapolated energy band-gap at 0 K, and the value found from the fit is $(3.240 \pm 0.005)$ eV. The other fit results $\alpha_{\rm V}$ and $\beta_{\rm V}$ used for the Varshni plot in Fig.~\ref{fig4}(c) are $0.022$ eV K$^{-1}$ and $3.25 \times 10^{4}$ K, respectively. Note that due to parameter correlations the error bars are large for both.



\subsection{\label{subs_DE} Direct edge with Wannier-Mott excitonic states}

\begin{figure}[tb]
\centering
\includegraphics[width =\columnwidth, clip, trim=50 0 20 40] {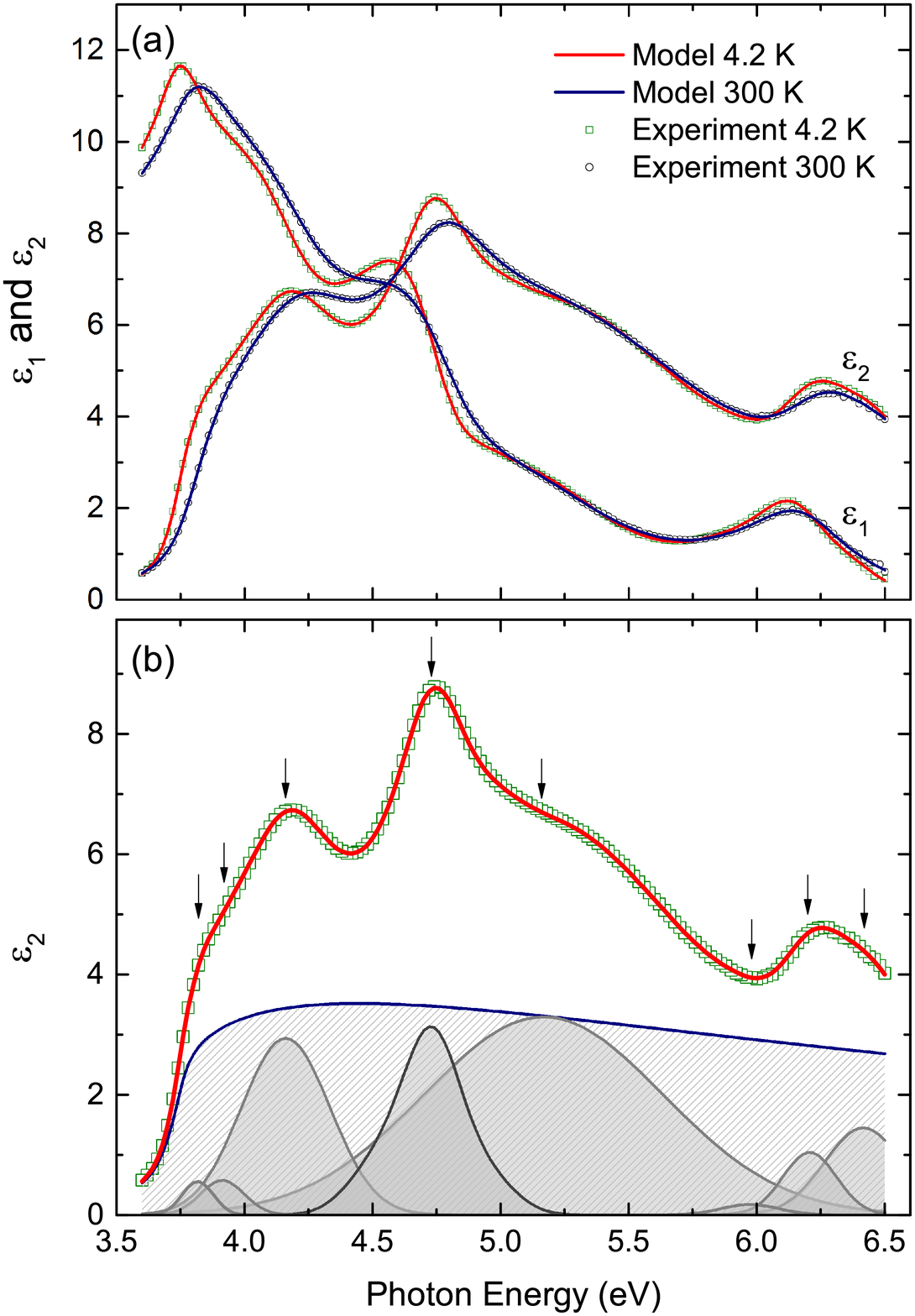}
\caption{\label{fig5}
Fit of the dielectric function  from above 3.6 eV using an excitonic band-edge term and other line shapes. (a)  Data (symbols) and fit (solid lines) of real and imaginary parts of the dielectric function at the lowest (4.2 K) and highest (300 K) measured temperatures as representatives. (b) The $\varepsilon_2$ data and fit at 4.2 K along with the individual line shapes. The arrows denote the center energies $E_n$ of the respective line shapes.}
\end{figure}

Direct edges in the optical spectra arise from dipole allowed transitions between equal momentum states \cite{YuCardona, Johnson1967}. A steep rise in the case of the fundamental edge  and  structures (e.g., peaks) for the subsequent edges are characteristics of direct transitions. The fundamental direct edge is often a distinct feature of the absorption spectrum of a material and its position can be identified qualitatively as the most prominent first  steplike feature.

The position of the fundamental direct edge of STO in all previous cases has been estimated using the one-electron (noninteracting) picture \cite{ vanBenthem, YuCardona, Pankove, Johnson1967}. In this one-electron picture, typically a direct edge is estimated using a linear fit of $\alpha^2E^2$ vs $E$, where $E$ is the photon energy \cite{YuCardona, Pankove, Johnson1967}. The intercept of such a linear fit with the energy axis gives the direct gap. Using the one-electron approach, the direct gap at 4.2 K and 300 K can be estimated to be 3.707 eV and 3.779 eV, respectively.


However, as previously revealed by the temperature-dependent pseudodielectric function with the support of \textit{ab initio} studies, excitonic effects play an important role in the optical spectra of STO \cite{Gogoi, Sponza}. It is noteworthy that otherwise in the past, excitonic effects have been ignored altogether in theoretical and experimental studies of optical spectra of STO. But, particularly with the availability of \textit{ab initio} methods based on Bethe-Salpeter equation (BSE) calculations on top of \emph{GW} results, it has been shown that in fact many-body interactions in terms of excitonic effects are indispensable for understanding the dielectric function of STO \cite{Sponza}. One of the main reasons for ignoring excitonic effects in STO is that its static dielectric constant is very high  at room temperature ($\sim$ 300) and even a few orders of magnitude higher at lower temperatures ($\sim$18 000 at 1.4~K~\cite{weaver}). In the case of such high static dielectric constants, usually excitonic effects are not expected or typically are negligible~\cite{dexter}.

In this scenario, excitonic effects in STO, despite its high dielectric constant,  render a new analysis of its band-edge characteristics \footnote{Studies have shown that $e$-$h$ interactions also play an important role in the manifested optical spectra of metals \cite{Mahan, Marini}, which in principle have an infinite static dielectric constant. In fact, excitons, which are screened on very fast time scales have been observed in silver recently \cite{Cui}.}. Therefore, we have performed detailed analysis of the fundamental direct optical absorption edge of STO for different temperatures, taking into considerations $e$-$h$ interactions. Elliott's formula  for the dispersion of Wannier-Mott excitons  gives an absorption coefficient including all bound and unbound excitonic states \cite{haug, Elliott}:
\begin{equation} \label{elliott}
\alpha(E) = \alpha_0\frac{E}{E_{\rm c}}\bigg[\sum_{n=1}^{\infty} \frac {4\pi}{n^3}\delta(\Delta+ \frac{1}{n^2}) + \Theta(\Delta)\frac{\pi e^{\frac{\pi}{\sqrt{\Delta}}}}{\sinh(\frac{\pi}{\sqrt{\Delta}})}\bigg],
\end{equation}

\noindent where $\Theta(\Delta)$ is the Heaviside unit-step function, $\Delta = (E-E_{\rm g})/E_c$, $E_{\rm c} = \hbar^2/(2m_ra_0^2)$, $a_0 = \hbar^2\varepsilon_0/(e^2m_r)$, $E_{\rm g}$ is the energy gap, $m_r$ is the $e$-$h$ reduced mass, and $\varepsilon_0$ is the static dielectric constant. $\alpha_0$ is a constant proportional to the square of the optical dipole matrix element~\cite{haug}, and $\delta$ represents the Dirac $\delta$ function.  The summation term on the right hand side  represents the bound states with sharp lines with rapidly decreasing intensity ($\propto 1/n^3$), while the second term represents the continuum of ionized states. The ratio of this continuum term together with the free carrier absorption, which is representative of the noninteracting case, gives the Coulomb enhancement factor,  $C(E) = \alpha_{\rm cont}/\alpha_{\rm free}$ \cite{haug}. In the presence of Coulomb interactions, particularly $e$-$h$ interactions, typically both bound excitons and an enhancement of the continuum state are found. However, since the observation of the bound states depends on the ratio of the binding energy to the broadening, it is possible that for a large broadening only an enhancement of the continuum states is seen~\cite{Tanguy3, saba2014}. As explained hereafter, in the case of STO, we observe no prominent bound states but considerable Coulomb enhancement of the continuum.

Particularly, we use an extended formulation  of Elliott's formula to include both the real and imaginary parts of the dielectric function in our analysis \cite{Goni, Tanguy1, Tanguy2, Tanguy3, Holden, Jellison_ZnO}. Including both the Lorentzian broadened  bound and unbound states, the total dielectric function near the band edge can be written as \cite{Tanguy1}
\begin{subequations} \label{Tanguyeq}
\begin{align}
\begin{split}
\varepsilon(E) = \frac{A\sqrt{E_{\rm b}}}{(E+i\gamma)^2}\{g(\xi(E+ i\gamma))
   + g(\xi(-E- i\gamma))\\-2g(\xi(0))\},  \label{eq:Tanguyeq1}\\
\end{split}
\intertext{where}
 & g(\xi) = 2\ln\xi - 2\pi \cot(\pi\xi) - 2 \psi(\xi) -\frac{1}{\xi}, \label{eq:Tanguyeq2}\\
 & \xi(z) = \sqrt{\frac{E_{\rm b}}{E_{\rm g} - z}}, \label{eq:Tanguyeq3}
\end{align}
\end{subequations}

\noindent and $\psi(z) = d\ln\Gamma(z)/dz$ \label{Tanguy3eq} is the digamma function. Here, $E_{\rm b}$ is the binding energy of the exciton, $\gamma$ is the broadening, and $A$ is a constant proportional to the square of the optical dipole matrix element~\cite{Tanguy1, Tanguy2, Tanguy3}.

\begin{center}
\begin{table*}[tbp]
\caption{Parameters of the different  line shapes for the fit of $\varepsilon$  at 4.2 K  and between 3.6 eV and 6.5 eV.} \label{tab2}
\renewcommand{\arraystretch}{1.7}
\begin{center}
\begin{tabular*} {0.9\linewidth}{@{\extracolsep{\fill}}cclll} \hline\hline \\[-18pt]
\textbf{No.} & \textbf{Description} & \multicolumn{3}{c}{\textbf{Parameters}} \\
\hline
1 &	$\varepsilon_1$ offset	& \multicolumn{3}{c}{2.019(17)}\\
2&	Excitonic band edge &  \multicolumn{3}{c}{\makecell{$A$ (eV$^2$) = 53.12(1.84), $E_{\rm g}$ (eV)=3.780(20), $\gamma$ (eV)=0.070(6),\\	
                                                                                                         $E_{\rm b}$ (eV)=0.022(8)}}\\
3 &	Gaussian &	$A_{\rm G}$ = 0.56(96) & $E_n$ (eV) = 3.817(19) & $\gamma$ (eV) =0.143(49) \\
4 &	Gaussian &	$A_{\rm G}$ = 0.58(75) & $E_n$ (eV) = 3.913(100) & $\gamma$ (eV) =0.201(163) \\
5 &	Gaussian &	$A_{\rm G}$ = 2.94(19) & $E_n$ (eV) = 4.160(9) & $\gamma$ (eV) =0.408(62) \\
6 & \makecell{Parameterized semiconductor \\
   oscillator function} & \multicolumn{3}{c}{\makecell{$A_{\rm P}$ = 6.10(32), $E_n$ (eV)=4.737(4), $\gamma$ (eV) =0.102(2), \\
   $W_{\rm L}$ (eV)=0.411(83),  $W_{\rm R}$  (eV)=0.387(8), $A_{\rm L}$ =0.207(12), $A_{\rm R}$=0.159(7)}}\\																																														
7 &	Gaussian &	$A_{\rm G}$ = 3.29(7) & $E_n$ (eV) = 5.167(9) & $\gamma$ (eV) =1.101(20)\\
8  & 	Gaussian &	$A_{\rm G}$ = 0.17(4) & $E_n$ (eV) = 5.973(36) & $\gamma$ (eV) =0.263(46)\\
9  &	Gaussian	& $A_{\rm G}$ = 1.04(30) & $E_n$ (eV) = 6.206(8) & $\gamma$ (eV) =0.248(22)\\
10 &	Gaussian	& $A_{\rm G}$ = 1.45(12) & $E_n$ (eV) = 6.416(25) & $\gamma$ (eV) =0.357(42)\\
\hline
\end{tabular*}
\end{center}
\end{table*}
\end{center}

For the direct band-edge term, Eq.~(\ref{Tanguyeq}) is used in the fitting routine for the dielectric function. At the same time, an asymmetric parametric semiconductor line shape \cite{Herzinger1, johs, herzinger2} and several  Gaussian line shapes with amplitudes $A_{\rm G}$, energies $E_n$, and broadenings $\gamma$ are used at higher energies for the fit~\cite{tompkins}. All variable fit parameters are as listed in Table~\ref{tab2}, together with  the results for the  dielectric function at 4.2 K. The experimental data have  been fitted only from above 3.6 eV to exclude the influence from the indirect edge.

The generated and the experimental dielectric functions for 4.2 K and 300 K  are plotted in Fig.~\ref{fig5}(a). Contributions of the individual oscillators and hence the excitonic band-edge line shape to the imaginary part of the dielectric function can be seen in Fig.~\ref{fig5}(b). The  features in the dielectric function above the fundamental direct edge can be represented well by single oscillators, except for the cases near 3.9 eV and 6.3 eV, where multiple adjacent oscillators  are required for an excellent agreement between experiment and fit. The structures starting from the direct edge until around 3.9 eV are likely due to transitions from the spin-orbit split-off bands of the highest valence band to the similarly split bands of the lowest conduction band. Such spin-orbit splittings have been predicted in recent studies, where the energy differences of the nondegenerate bands are in the range of a few tens of millielectronvolts, which is similar to the observed differences of the oscillator center energies in our case \cite{Marques, sahin}. The exact determination of the critical point energies is necessary for precise assignment of such transitions.  On the other hand, the multiple oscillator composition of the dielectric function around 6.3 eV could be due to more subtle effects, which need further investigation.


\begin{figure}[tb]
\centering
\includegraphics[width =\columnwidth, clip, trim=55 40 15 35] {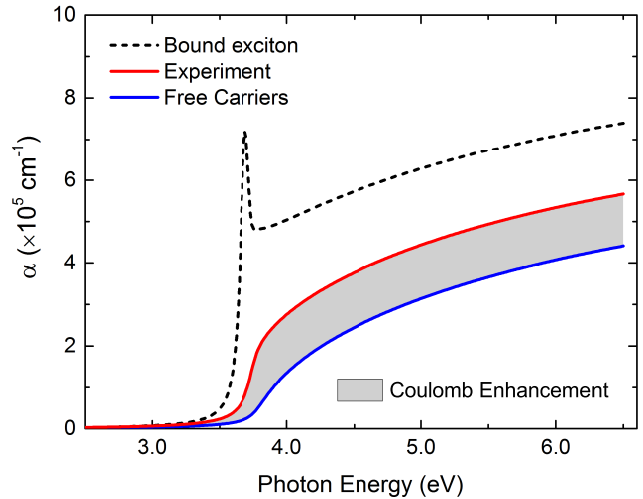}
\caption{\label{fig6}
Comparison of the experimental band-edge absorption  with the noninteracting free carrier absorption. The shaded region represents the Coulomb enhancement due to interactions.}
\end{figure}

The fitting results from the excitonic  band-edge term contain the important values regarding the fundamental direct band-gap,  binding energy of the excitonic interaction, and the Lorentzian broadening.  The experimental excitonic band-edge line shape  using the results of the fit is depicted in Fig.~\ref{fig6}.  The free carrier absorption, representing the noninteracting case, is estimated by setting  the binding energy to zero and  keeping other parameters unchanged. It should be noted that the expression for the absorption in the noninteracting case of  a three-dimensional M$_0$ critical point gives the same result~\cite{YuCardona, Mcardona}. The Coulomb enhancement can now be estimated, and it is  found to be  2.05 at 4 eV and 1.31 at 6 eV, for example \cite{haug, Tanguy1}. In Fig.~\ref{fig6}, the Coulomb enhancement region is plotted  for the case of 4.2 K.  As an example, the case of a bound exciton is also shown  and is obtained by using a binding energy of 100 meV and broadening of 40 meV instead of the respective fit results, with other terms kept the same. However, the ratio of the binding energy to the broadening value is small for STO, and hence no bound states exist~\cite{Tanguy3, saba2014}. The broadening term determines the intensity and shape of the bound states as well as for the unbound region, e.g., with higher broadening the peak intensity of the bound exciton  will decrease.

\begin{figure}[tb]
\centering
\includegraphics[width =\columnwidth, clip, trim=0 0 0 0] {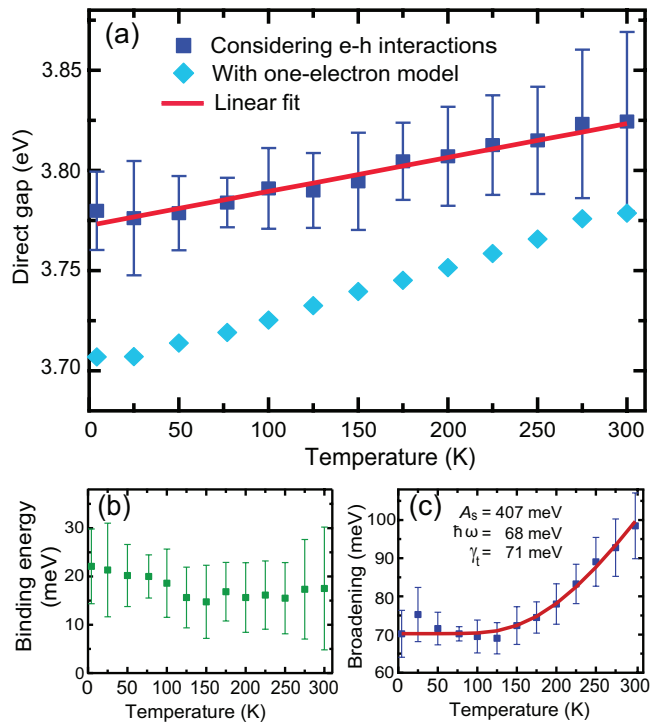}
\caption{\label{fig7}
Extracted fundamental direct gap of STO and its temperature dependence. Temperature dependence of (a) the fundamental direct gap of STO and comparison with the one- electron gap, (b) the excitonic   binding energy, and (c) the Lorentzian broadening of the interband transition associated with the excitonic band edge.}
\end{figure}

The determined fundamental direct gaps, binding energies, and broadenings are plotted for all the measured temperatures in Fig.~\ref{fig7}. Additionally, the one-electron band gap is also plotted in Fig.~\ref{fig7}(a) for comparison. As can be seen, the gap obtained considering excitonic interactions differs from the one-electron gap by about 73 meV at 4.2 K, and this difference decreases slightly with increasing temperature. It is remarkable that an almost  linear increase is observed for the fundamental direct edge with the increase of temperature, as shown in Fig.~\ref{fig7}(a). This is in contrast to the indirect edge, which follows Varshni's equation in the cubic phase. Moreover, it should be noted that with the increase of temperature,  band-gaps of most insulators and semiconductors decrease with a typical coefficient of $-3$~$\times$~$10^{-4}$ to $-5$~$\times$~$10^{-4}$~{\rm eV}/{\rm K}~\cite{Mcardona}. A notable exception is the temperature dependence of the fundamental direct edge of the lead chalcogenides, which show a positive coefficient of about 4~$\times$~$10^{-4}$~{\rm eV}/{\rm K}~\cite{scanlon}. In the case of STO, we find a similar anomalous positive coefficient of 1.7~$\times$~$10^{-4}$~{\rm eV}/{\rm K} from the linear fit in Fig.~\ref{fig7}(a). In general, the phenomenon of a positive band gap change with increasing temperature is believed to be due to the presence of hybridized $p$-$d$ orbitals~\cite{CardonaMkremer}. Our results of STO, which also exhibits $p$-$d$-electron hybridization, further support this hypothesis.



On the other hand, the binding energy, based on Elliott's formula for the dispersion of Wannier-Mott excitons and plotted in Fig.~\ref{fig7}(b), remains almost constant at around 20 meV for the full experimental temperature range. Note that the exciton binding energy is found to be lower than what has been reported previously~\cite{Sponza, Gogoi}. In general, it can be noted that the binding energy is in the range of the typical Wannier-Mott exciton binding energy \cite{YuCardona}, and it is still substantial in strength considering the large static dielectric constant of STO. Particularly, it is not inversely proportional to the square of the static dielectric constant  and also does not reduce with decreasing temperature, even though the static dielectric constant increases steeply \cite{dexter}.

The  temperature dependence of the broadening remains almost constant until  around 125 K and then starts to rise linearly [Fig.~\ref{fig7}(c)]. This potentially  indicates that the broadening depends on the microstructure, and the phase transition at 105 K from tetragonal to cubic crystal structure plays a role here. Furthermore, the excitonic broadening follows Toyozawa's equation given by~\cite{Jeong, Toyozawa1, Toyozawa2} 
\begin{equation} \label{Toyozawa}
\gamma = \frac{A_{\rm s}}{e^{{\hbar\omega}/kT} -1} + \gamma_{\rm t}.
\end{equation}
 
\noindent Here, the constants $A_{\rm s}$, $\gamma_{\rm t}$, and $\hbar\omega$ can be extracted from the fit. The constant $\gamma_{\rm t}$ represents the threshold broadening at low temperatures and is much higher than the binding energy. The determined average energy of the participating phonons $\hbar\omega=(68\pm25)$ meV is close to the reported LO$_2$ phonon energy of 58 meV~\cite{Servoin}.

\section{\label{concl} Conclusion}

In summary, temperature-dependent spectroscopic ellipsometry measurements have been performed  over a wide energy range to obtain the complex  dielectric functions which reveal important features of the electronic structure.  The lowest temperature dielectric function at 4.2 K presented here is suitable  for comparison with theoretical results,  where usually lattice vibrations are neglected.  The temperature-dependent dielectric functions are analyzed with regard to  state-of-the-art theoretical calculation results, eliminating existing ambiguities in the interpretations. Particularly, the fundamental indirect and direct band edges are investigated for the full temperature range. It is found that the temperature dependence of the indirect edge follows Varshni's rule above the phase transition temperature (105 K). The direct edge is analyzed using an excitonic band-edge line shape. Notably, it is found to be different from the one-electron gap. The temperature dependence of the fundamental direct gap shows an anomalous behavior with an almost linear increase with increasing temperature. An important  observation is that the structural phase transition at 105 K affects the evolution of the fundamental indirect edge as well as the Urbach tail conspicuously, while the position of the fundamental direct edge is unaffected.   

Overall, these results present an updated, deeper, and coherent insight into the electronic structure of STO,  which is crucial for understanding its properties. This understanding has direct relevance in elucidating the novel fundamental physical phenomena  of STO heterostructures.

\begin{acknowledgments}
We would like to thank A. Rusydi for a critical reading of the manuscript. This work is supported by the Singapore National Research Foundation under its Competitive Research Funding (NRF-CRP 8-2011-06 and NRF2008NRF-CRP002024) and MOE-AcRF Tier-2(MOE2010-T2-2-121).
\end{acknowledgments}

%

\bibliographystyle{apsrev4-1}
\bibliography{TDSTO}

\end{document}